\documentstyle[prl,aps,multicol]{revtex}

\begin{document}

\preprint{Preprint}
\draft{}

\title{Self-organized criticality in the Bean state in
YBa$_2$Cu$_3$O$_{7-x}$ thin films.}

\author{C.~M.~Aegerter, M.~S.~Welling, and R.~J.~Wijngaarden}

\address{Division of Physics and Astronomy, Faculty of Sciences,
Vrije Universiteit, De Boelelaan 1081, 1081HV Amsterdam, The Netherlands}

\date{\today}
\maketitle

\begin{abstract}
The penetration of magnetic flux into a thin film of
YBa$_2$Cu$_3$O$_{7-x}$ is studied when the external field is
ramped slowly. In this case the flux penetrates in bursts or
avalanches. The size of these avalanches is distributed according
to a power law with an exponent of $\tau$ = 1.29(2). The
additional observation of finite-size scaling of the avalanche
distributions, with an avalanche dimension D = 1.89(3), gives
strong indications towards self-organized criticality in this
system. Furthermore we determine exponents governing the
roughening dynamics of the flux surface using some universal
scaling relations. These exponents are compared to those obtained
from a standard roughening analysis.
\end{abstract}

\begin{multicols}{2}
The critical state in a type-II superconductor shows
a powerful analogy to a granular pile, which was already noted by
de Gennes in the 1960s \cite{degennes}. With the advent of
self-organized criticality (SOC) \cite{SOC}, the avalanche
behavior of granular piles was intensely studied \cite{sand}. This
is because SOC was thought of as a general mechanism to explain
the intermittent behavior of slowly driven systems far from
thermodynamic equilibrium. The experimental verification of SOC,
however, was not straightforward. Power-law distribution of
avalanches was not observed in many experiments \cite{jaeger}. One
of the hall-marks of critical behavior, finite-size scaling, was
only found in a few cases, most notably in experiments on a one
dimensional (1d) pile of rice \cite{Frette}, as well as in a 1d
pile of steel balls with a random distribution of balls in the
bottom layer \cite{ernesto2}. The only study of finite size
scaling in a 2d system to date, is to our knowledge a study by
some of us on the properties of a 2d pile of rice \cite{ricepile},
which is qualitatively similar to the one presented here.

Given the classical analogy with granular piles \cite{ernesto},
the critical state in superconductors was also quickly proposed as
a SOC system \cite{tang}. Experimentally, magnetic vortices are
well suited to study SOC, since kinetic effects, which can lead to
deviations from critical behavior in sand-piles
\cite{Frette,feder}, are naturally suppressed due to their
overdamped dynamics \cite{blatterbible}. Just as in the case of
granular piles, however, the experimental confirmation of this
conjecture has been controversial. While power-law behavior in the
avalanche distribution has been observed by most authors
\cite{field,behnia}, finite-size scaling was not observed so far
in the case of the critical state in superconductors. This is
because most of the studies \cite{field,aval}, were carried out
using magnetization measurements, which only give information on
the overall behavior of the whole of the sample. This corresponds
to only considering off-edge avalanches in a sand-pile, which do
not capture the full dynamics and may therefore give a flawed
picture \cite{feder}. More recent investigations using arrays of
miniature Hall-probes \cite{behnia}, do give insights into
internal avalanches and the complete dynamics, but only give
information from a few selected points in the sample, which makes
testing for finite-size scaling impossible.

In this letter, we study the local changes in the magnetic flux
over the whole central area of a sample. This is done via a highly
sensitive magneto-optic setup, which can resolve flux densities of
0.2 mT over an area of $\sim$ 5x5 $\mu$m$^2$. This implies that
flux changes corresponding to 2.5 $\Phi_0$ can be resolved, where
$\Phi_0$ = h/2e, is the magnetic flux quantum (the flux of a
single vortex). Furthermore, since we observe a large field of
view, different size subsets can be studied leading to the
possibility of observing finite-size scaling. Given the
finite-size scaling exponents, one can determine the roughening
and dynamical exponents of the flux-surface from universal scaling
relations. Comparing these results with those determined directly
from the surface properties \cite{marco}, gives good agreement.
Furthermore, the exponents fulfill the scaling relation $\alpha +z
=2$, valid for the well known driven interface model given by the
Kardar-Parisi-Zhang (KPZ) equation \cite{KPZ} (see below).

The experiments were carried out on a thin film of
YBa$_2$Cu$_3$O$_{7-x}$ (YBCO), grown on a NdGaO$_3$ substrate to a
thickness of 80 nm using pulsed laser ablation \cite{sample}. The
pinning sites in the sample are uniformly distributed and consist
mostly of screw dislocations acting as point pins \cite{defects}.
A polarization microscope was placed in an Oxford Instruments
cryomagnet capable of a maximum field of 1 T and cooling to a
temperature of 1.8 K. The sample was cooled in zero field to 4.2
K, at which point the external field was slowly increased in steps
of $\mu_0\Delta H$ = 50 $\mu$T, after which the sample was allowed
to relax for 10 s before an image was taken. In a magneto-optical
experiment, the flux density $B_z$(x,y) at the surface of the
sample is measured via an indicator layer showing a strong Faraday
effect. The polarization of the incoming light is turned in
proportion to the flux density in the indicator layer. A
cross-polarized analyzer will thus admit light from regions with
non-zero magnetic flux density \cite{huebener}. However, in order
to determine the rotation angle (and thus the flux density B$_z$)
directly, including its sign, we employ a recently developed
magneto-optic image lock-in amplifier \cite{arvoo}, using a
modulation of the incoming polarization vector. The output of the
instrument gives directly the Faraday angle for each pixel,
independent of spatial inhomogeneities in the illumination.

\begin{figure}
\input{epsf}
\epsfxsize 8cm 
\centerline{\epsfbox{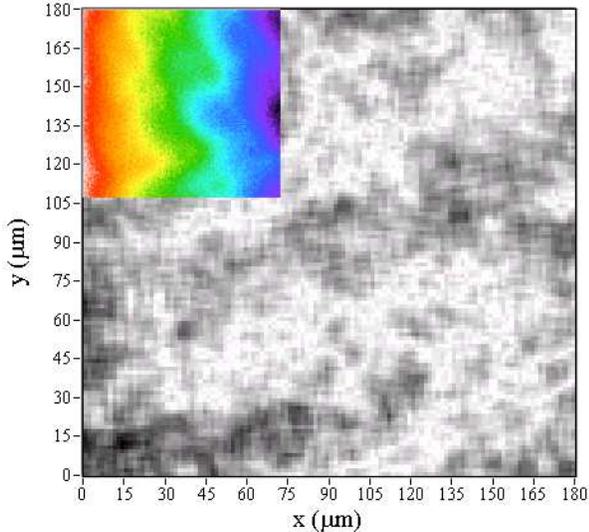}} \caption[~]{Snapshot of a
flux-avalanche. Shown is the difference between two consecutive
field-profiles $\Delta B (x,y)$. The scale goes from 0 (white) to
0.4 mT (black). The area-integral over this difference corresponds
to the size of the flux-jump $\Delta\Phi$. In order to determine
the roughness exponent of the whole surface, we also need to
determine the fractal dimension of the active sites in such an
avalanche (see text). The inset shows the flux distribution
$B_z$(x,y) in the same part of the sample. Here the scale varies
from 0mT (black) to 12mT (white). Lines of constant field can be
identified by the different shades.} \label{diff}
\end{figure}

\begin{figure}
\input{epsf}
\epsfxsize 9cm 
\centerline{\epsfbox{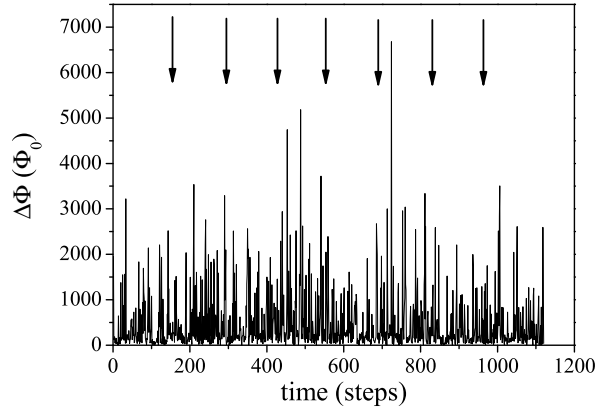}} \caption[~]{The time evolution of
the magnetic flux inside the sample over all nine experiments. The
magnetic field difference $\Delta B$ has been integrated over the
whole surface of 180x180 $\mu$m$^2$ and the average flux increase
has been subtracted. The evolution takes place in the form of flux
jumps or avalanches of various sizes, which are summarized in the
histogram of Fig.~\ref{hist}. The different experimental runs are
separated by arrows.} \label{time}
\end{figure}

The data analyzed here come from a series of nine experimental
runs, each consisting of $\sim$ 300 time-steps. Of these steps
only the last 140 in each run were used, in order to have a
critical state established in the whole region of the sample used
for analysis. The size and shape of the avalanches was determined
from the difference $\Delta B_z$(x,y) of two consecutive images
(see Fig.~\ref{diff}). From this difference, the average increase
in the applied magnetic field, due to the step-wise field-sweep,
was subtracted in order to solely study the avalanches. The
external fields were determined from a region well outside the
sample, which was also in the field of view. Once the incremental
field difference is determined, the size of an avalanche,
corresponding to the displaced amount of flux $\Delta\Phi$, is
obtained from $\Delta B$ via integration over the whole area
\begin{equation}
s = \Delta\Phi = \frac{1}{2} \int |\Delta B(x,y)| dx dy.
\end{equation}
The average increase in the external field is in good agreement
with the increase in flux density of 50 $\mu$T between images. The
resulting time series of the avalanche behavior of all the
experiments is shown in Fig.~\ref{time}. In this figure, $\Delta
B$ was integrated over the whole area of 180x180 $\mu$m$^2$ in
each time step. As can be seen, the evolution of the magnetic flux
inside the sample is intermittent with occasional large jumps.

\begin{figure}
\input{epsf}
\epsfxsize 9cm 
\centerline{\epsfbox{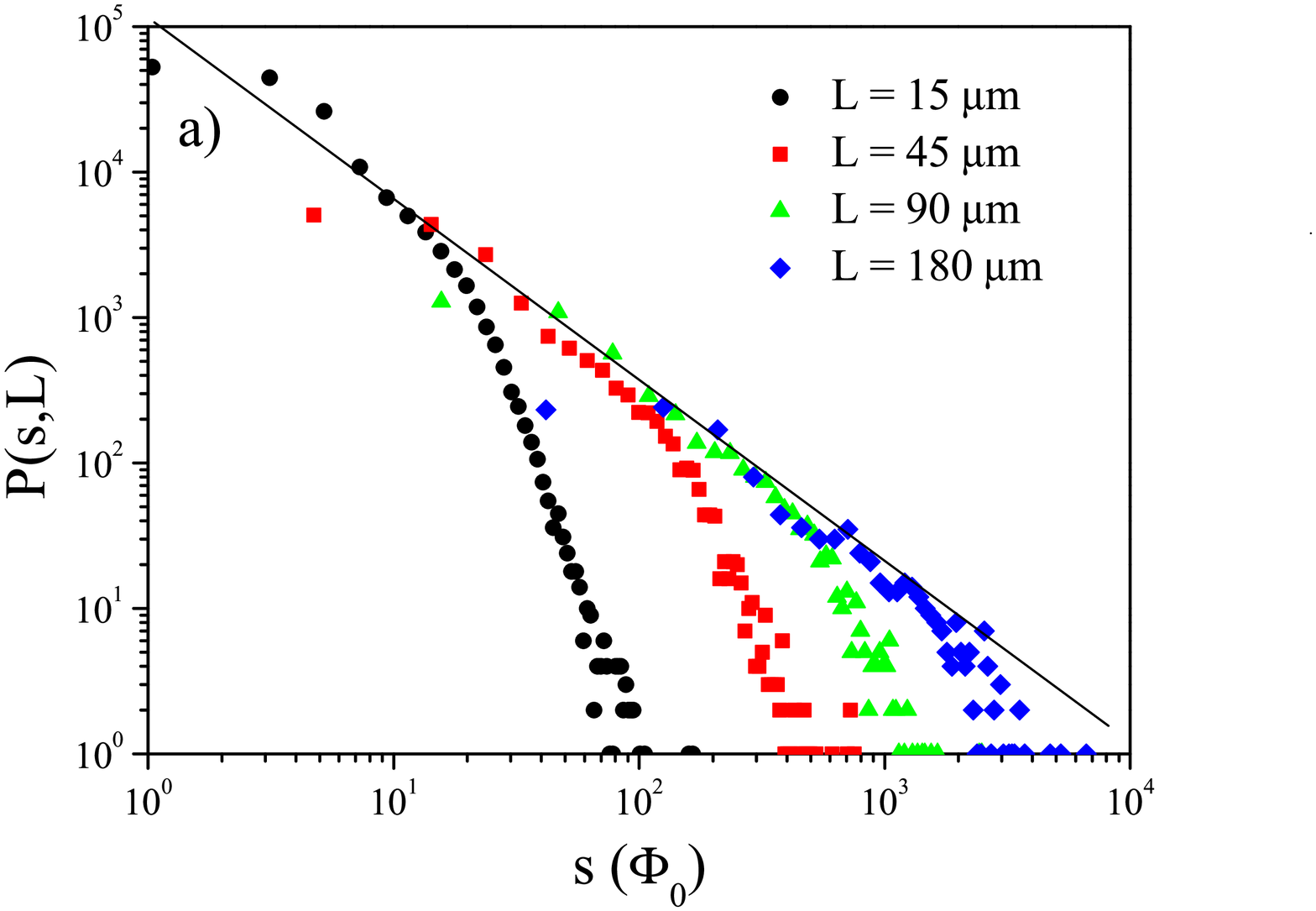}} \epsfxsize 9cm
\centerline{\epsfbox{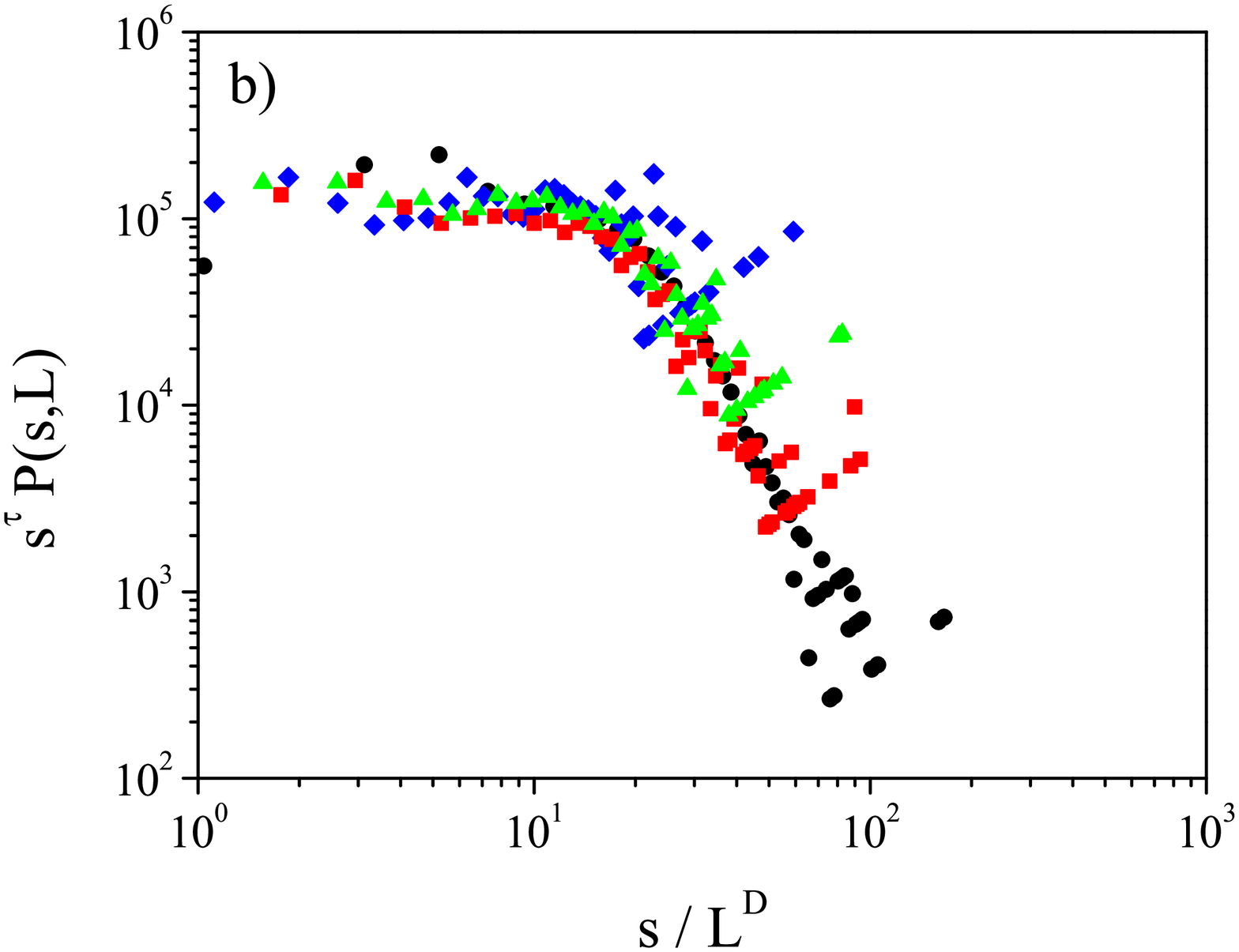}} \caption[~]{(a) The direct
avalanche size distribution for different sizes of windows (180,
90, 45, and 15 $\mu$m respectively). There is an envelope to the
distribution in the form of a power-law indicated by the black
line over more than three decades with an exponent of $\tau$ =
1.30(5). (b) The scaled size distributions showing a curve
collapse. The exponents used in the finite-size scaling are $\tau$
= 1.29(2) and $D$ = 1.89(3).} \label{hist}
\end{figure}

In order to check the data for finite size scaling, we also
integrated $\Delta B$ over subsets of the image of a linear size
of $L$ = 90, 45 and 15 $\mu$m respectively. The histograms of the
avalanche size distribution for these data-sets are shown unscaled
in Fig.~\ref{hist}a. As can be seen, the smallest avalanches
correspond to a flux change of 2-3 $\Phi_0$, corresponding to the
resolution of the measurement. Taking all of the data together we
observe a power-law distribution over more than three decades. The
slope of the black line gives the exponent of the distribution,
$\tau$ = 1.30(5). In Fig.~\ref{hist}b we show the same data, but
now the probabilities are scaled with $s^\tau$ and the avalanche
sizes are scaled with $L^{-D}$. As can be seen, there is very good
curve collapse indicating the presence of finite size scaling
\cite{barabasi}. This means that the avalanche size distribution
function is given by
\begin{equation}
P(s,L) = s^{-\tau} f(\frac{s}{L^D}),
\end{equation}
where $f(x)$ is constant up to a cutoff scale $s_{co} \propto
L^D$. The values of the exponents used to obtain curve collapse
are $\tau$ = 1.29(2) and $D$ = 1.89(3). Note that here, we have
carried out the finite size scaling by way of subdividing the
whole image rather than carrying out experiments with different
size samples. We have checked by means of simulations of the 2d
Oslo-model that such a subdivision into finite-size samples leads
to the same scaling exponents as a curve collapse of different
simulations of finite size. Furthermore, an independent
measurement of $D$, directly using a box counting method
\cite{mandelbrot} in 3d yields $D = 1.92(5)$, consistent with the
value from finite size scaling.

It has already been noted \cite{radu} that the front of
penetrating flux shows kinetic roughening. Similarly, the surface
$B_z(x,y)$ in two dimensions can be shown to be self-affine, with
a roughness exponent $\alpha$, characterizing the growth of the
interface width $w_{sat}$ with the system size, $w_{sat} \propto
L^\alpha$ and a dynamic exponent $z$ characterizing the saturation
time of the width. In such an analysis of the present data,
discussed elsewhere \cite{marco}, the roughness exponent was found
to be $\alpha$ = 0.73(5) and the dynamic exponent was found to be
$z$ = 1.38(10). Using the universal scaling relations derived by
Paczuski {\em et al.} \cite{pacz} for various SOC models, these
exponents characterizing the roughness and growth of the surface
can also be determined from the scaling exponents of the avalanche
distribution. This indicates the fact that the roughening of the
surface originates from the avalanche distribution and its
underlying dynamics.

\begin{figure}
\input{epsf}
\epsfxsize 9cm 
\centerline{\epsfbox{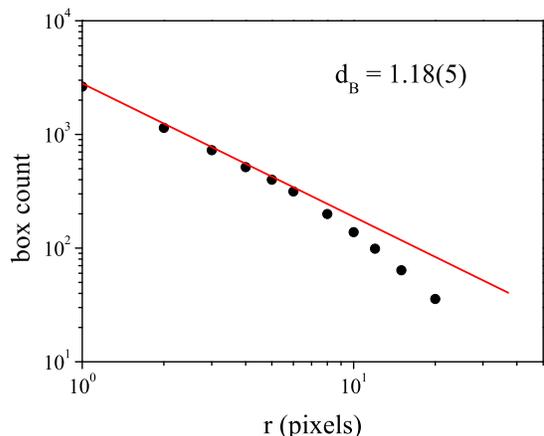}} \caption[~]{Determination of the
fractal dimension of the active clusters in an avalanche using the
box counting method. In this determination, avalanches more than
one standard deviation larger than the average size have been
studied. The analysis of one such cluster is shown. The fractal
dimension is given by the slope of the straight line. Averaged
over all clusters, this leads to a fractal dimension of the active
clusters of $d_B$ = 1.18(5). } \label{frac}
\end{figure}

Let us first discuss the roughness exponent. According to the
finite size scaling, an avalanche of the cut-off size will scale
like $s_{co} \propto L^D$. Similarly, the size of such an
avalanche will roughly be given by $s_{co} \simeq w_{sat}L^{d_B}$,
where $d_B$ is the fractal dimension of the area of an avalanche.
Equating the two expressions for $s_{co}$, one obtains $w_{sat}
\propto L^{D-d_B}$ and hence
\begin{equation}
\alpha = D - d_B, \label{alpha}
\end{equation}
in agreement with Ref.~\cite{pacz}. In order to determine the
roughness exponent from the avalanches, we therefore have to
measure the fractal dimension, $d_B$, of the avalanches area,
which was done using a simple box-counting method
\cite{mandelbrot}. To this end, avalanches which were one standard
deviations bigger than average were binarized, yielding a
distribution of active clusters used in the determination of the
fractal dimension. The result for one such cluster can be seen in
Fig.~\ref{frac}, where the number of active pixels of the
avalanche is shown as a function of the length scale. Averaged
over all clusters analyzed, the fractal dimension, given by the
slope of the line in Fig.~\ref{frac}, is $d_B$ = 1.18(5). From
this we determine the roughness exponent as $\alpha$ = 0.71(5), in
good agreement with that determined from a roughness analysis (via
the correlation function and the power-spectrum) of the surface
fluctuations \cite{marco}.

The dynamic exponent, $z$, can be obtained from a similar
argument. The scaling of the crossover-time with the system size
is described by $z$, $t_\times \propto L^z$. This crossover-time
can be roughly estimated from the time it takes for an avalanche
of the cut-off size $s_{co}$ to appear. Since flux is added to the
system in constant steps, $\delta\Phi$, the number of vortices
added until a cut-off avalanche occurs is given by $\delta\Phi
t_\times$. On the other hand, the flux added will also be moved in
smaller avalanches. The total flux necessary to be introduced in
order to obtain an avalanche of size $s_{co}$ can be estimated
from integrating over all avalanche sizes up to the one of size
$s_{co}$:
\begin{equation}
t_\times \propto \int_0^{s_{co}} s P(s) ds \propto
s_{co}^{2-\tau}.
\end{equation}
Using $s_{co} \propto L^D$, we obtain $t_\times \propto L^{D(2 -
\tau)}$,which immediately leads to the scaling relation
\begin{equation}
z = D(2-\tau). \label{z}
\end{equation}
Again, this is also in agreement with the universal scaling
relation derived by Paczuski {\em et al.} \cite{pacz}. Inserting
the values determined above, we obtain $z$ = 1.34(4), again in
good agreement with that determined via roughness analysis
\cite{marco}.

In conclusion, we have shown that the distribution of the size of
flux jumps in an YBCO thin film is not only given by a
power-law,but also shows finite size scaling. Due to universal
scaling relations valid for a SOC state \cite{pacz}, the exponents
determined via finite-size scaling can also be used to describe
other properties of the system. One such example is the roughening
of the magnetic flux surface. Here the statistical properties of
the self-affine structure are built up by the penetrating flux,
such as the roughness and dynamical exponent can be derived from
the structure and dynamics of the flux-avalanches using relations
(\ref{alpha}) and (\ref{z}). These characteristic exponents can
however also be determined directly, via a roughness analysis
\cite{marco}, the results of which can be compared to those
obtained via the avalanche dynamics. As discussed above, we find
excellent agreement between the exponents determined in these
separate ways. The critical state observed in the YBCO thin films
can be seen as a realization of a 2d roughening system, albeit
with a self-organized dynamics. In this context, we note that the
roughness and dynamic exponents fulfill the general KPZ scaling
relation $\alpha + z = 2$, which is an exact result also in 2d.

In the future, the magnetic flux structures in superconductors, as
the ones studied here, may be used as an ideal experimental system
with which to study non-equilibrium phenomena, especially those of
granular matter. In fact we find strong qualitative correspondence
of the behavior of the vortices with that of a pile of rice
\cite{ricepile} in terms of SOC behavior. However, in addition to
the granular system, the superconducting system studied here
allows for the presence and control of quenched noise due to
pinning sites, whose influence on the statistical properties can
then be studied experimentally as well. Moreover, the physics of
the microscopic behavior of vortices is well studied
\cite{blatterbible}, such that collective effects can be studied
directly.

We would like to thank Jan Rector for the growth of the samples.
This work was supported by FOM (Stichting voor Fundamenteel
Onderzoek der Materie), which is financially supported by NWO
(Nederlandse Organisatie voor Wetenschappelijk Onderzoek).

\bibliographystyle{prsty}

\end{multicols}
\end{document}